\documentclass[aps,prl,reprint,twocolumn,superscriptaddress]{revtex4-1}
\usepackage{euscript}
\usepackage{amssymb}
\usepackage{amsfonts}
\usepackage{amsbsy}
\usepackage{epsfig}
\usepackage{amsthm}
\usepackage{amscd}
\usepackage{amstext}
\usepackage{verbatim}
\usepackage{amsmath}
\usepackage{cancel}
\usepackage{capt-of}
\usepackage{empheq}
\usepackage{subfigure}
\usepackage{xcolor}

\usepackage[pdftex]{hyperref}
\hypersetup{ 
colorlinks=true, 
linkcolor=blue, 
citecolor=red, 
}

\renewcommand{\(}{\left(}
\renewcommand{\)}{\right)}
\def\be{\begin{equation}}
\def\ee{\end{equation}}

\begin{document}
\frenchspacing

\title{Causality Constraints in Large $N$ QCD Coupled to Gravity}

\author{Jared Kaplan}
\email{jaredk@jhu.edu}

\author{Sandipan Kundu}
\email{kundu@jhu.edu}

\affiliation{\vspace*{4mm}
Department of Physics and Astronomy, 
Johns Hopkins University,
Baltimore, Maryland, U.S.A.}

\begin{abstract}

Confining gauge theories contain glueballs and mesons with arbitrary spin, and these particles become metastable at large $N$. However, metastable higher spin particles, when coupled to gravity, are in conflict with causality. This tension  can be avoided only if the gravitational interaction is accompanied by interactions involving other higher spin states well below the Planck scale $M_{\rm pl}$. These higher spin states can come from either the QCD sector or the gravity sector, but both these resolutions have some surprising implications. For example, QCD states can resolve the problem since there is a non-trivial mixing between the QCD sector and the gravity sector, requiring all particles to interact with glueballs at tree-level. If gravity sector states  restore causality, any weakly coupled UV completion of the gravity-sector  must have many stringy features, with an upper bound on the string scale. Under the assumption that gravity is weakly coupled, both scenarios imply that the theory has a stringy description above $N\gtrsim \frac{M_{\rm pl}}{\Lambda_{\rm QCD}}$, where $\Lambda_{\rm QCD}$ is the confinement scale.

\end{abstract}

\maketitle

\noindent {\it \bf Introduction.---} 
In the early days of quantum chromodynamics (QCD), 't Hooft pointed out that there is an unconventional systematic expansion obtained by taking the number of colors $N\rightarrow \infty$ and the gauge coupling $g\rightarrow 0$ with 't Hooft coupling $\lambda=g^2 N=$ fixed \cite{tHooft:1973alw, tHooft:1974pnl}. 
 Any such confining gauge theory is characterized by  a confinement scale $\Lambda_{\rm QCD}$ where the 't Hooft coupling becomes strong. This scale determines the characteristic mass and physical size of generic hadrons bound together by the confining force.
 
  It has been long believed that  the confinement persists even at large $N$. This assumption leads to a classic result about the scaling of the correlators or scattering amplitudes of mesons ($\pi$) and glueballs ($G$) in the large $N$ limit (in the absence of gravity)
\cite{tHooft:1973alw, tHooft:1974pnl,Witten:1979kh}
\begin{align}\label{eq:GGG}
\langle G_1 \cdots G_n\pi_1 \cdots \pi_p \rangle \sim \frac{1}{N^{n+\frac{p}{2}-1-\delta_{p,0}}}\ ,
\end{align}
where amplitudes for free propagation are normalized so that they are independent of $N$. Therefore,  the lifetime of a meson is rather long $\sim O(N)$. Glueballs are even more stable with typical lifetime of order $O(N^2)$.  Furthermore, it is expected that QCD contains glueballs and mesons of spin $J>2$. On physical grounds, this is obvious since it is possible to construct color singlet states with arbitrary spin by spinning quarks and gluons. The scaling relation (\ref{eq:GGG}) immediately implies that   in the exact $N=\infty$ limit these higher spin mesons and glueballs behave as stable particles that are free and non-interacting. 

However, gravitational interactions of higher spin particles are strongly constrained by causality \cite{Afkhami-Jeddi:2018apj, Kaplan:2019soo, Kaplan:2020ldi}. In particular, a recent theorem on metastable higher spin particles seems to suggest that glueballs or mesons of spin $J>2$ when coupled to gravity can be used to send signals outside of the lightcone \cite{Kaplan:2020ldi}. In this Letter, we discuss how this  tension between causality and confining large $N$ gauge theories in $3+1$ dimensions can be resolved.  We discuss two classes of resolutions: (I) gauge theory states might resolve this tension via a mixing with gravity or (II) new gravity sector states could remove the problem.  Scenario (I) is certainly the more conservative resolution, as it does not involve any new states. However, for it to work a tower of spin-2 glueballs must remove causality violation due to pure graviton exchange between higher spin glueballs.  We believe this possibility merits further investigation, since the second scenario would have profound implications.

{\it \bf Causality.---} The theorem of \cite{Kaplan:2020ldi} leverages recent advances  constraining quantum field theories (QFTs) from causality \cite{Hofman:2008ar,Hofman:2009ug,Camanho:2014apa,Hartman:2015lfa,Hartman:2016dxc,Edelstein:2016nml,Hofman:2016awc,Hartman:2016lgu,Afkhami-Jeddi:2016ntf,Camanho:2016opx,Afkhami-Jeddi:2017rmx,Hinterbichler:2017qcl, Bonifacio:2017nnt, Afkhami-Jeddi:2018own,Afkhami-Jeddi:2018apj,Kaplan:2019soo,Chowdhury:2018nfv,Kundu:2020gkz}. The main idea parallels the philosophy of \cite{Adams:2006sv} which  demonstrated that there are Wilsonian effective field theories without a consistent UV completion.

\begin{figure}[h]
\begin{center}
\includegraphics[width=0.35\textwidth]{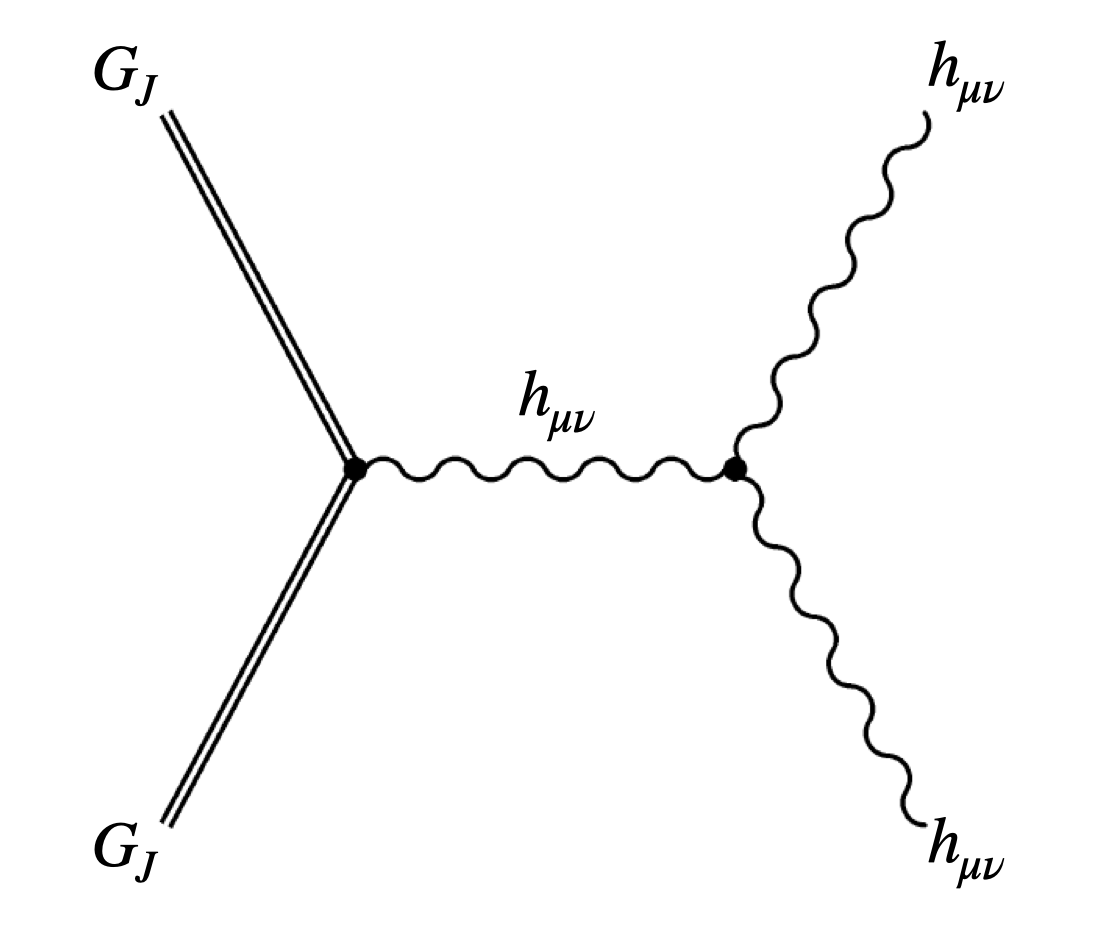}
\end{center}
\caption{ \small This Feynman diagram in large $N$ QCD is inconsistent with causality for $J\ge 3$ unless it is accompanied by an infinite tower of additional higher spin exchanges  with unbounded spin. The same is true for mesons as well. \label{fig:feynman}}
\end{figure}

In QFT, the eikonal phase-shift $\delta(s,\vec{b})$, where $\vec{b}$ is the impact parameter, plays a crucial role since it is closely related to the Shapiro time-delay. In particular, causality requires that (i) $\delta(s,\vec{b})$ cannot grow faster than $s$, (ii) when $\delta(s,\vec{b})$ grows as $s$ it must be non-negative \cite{Camanho:2014apa}. This imposes constraints on the scattering amplitude $\mathcal{A}(s,t)$ of the $2\rightarrow 2$ scattering $G_J h_{\mu\nu}\rightarrow G_J h_{\mu\nu}$, where $h_{\mu\nu}$ is the graviton and $G_J$ is any glueball in large $N$ QCD with spin $J\ge 3$. It is a well-known fact that only t-channel poles of $\mathcal{A}(s,t)$ contribute to the phase-shift. For weakly coupled gravity, $\mathcal{A}(s,t)$ is a meromorphic function with simple poles only at locations corresponding to particles in the theory. Thus, $\mathcal{A}(s,t)$ must have a simple pole at $t=0$ which represents a graviton exchange. However, the contribution of the graviton pole to the phase-shift is inconsistent with the positivity condition $\delta(s,\vec{b})\ge 0$ because of interference effects \cite{Afkhami-Jeddi:2018apj,Kaplan:2020ldi}. This implies that the Feynman diagram \ref{fig:feynman} for $J\ge 3$, by itself, is at odds with causality.

The bounds from \cite{Kaplan:2020ldi}, when applied to glueballs of large $N$ QCD, imply that this causality violation can only be avoided if the scattering amplitude $\mathcal{A}(s,t)$ has other t-channel poles that correspond to higher spin ($J\ge 3$) states with an upper bound on the the mass $\Lambda_{\rm HS}$ of the lightest higher spin t-channel pole  \cite{comment_meson}:
\begin{align}\label{eq:BoundOnStringScale}
&\Lambda_{\rm HS} \lesssim \Lambda_{\rm QCD}\(\frac{M_{\rm pl}}{N \Lambda_{\rm QCD}}\)^{\gamma} \qquad  N\lesssim \frac{M_{\rm pl}}{\Lambda_{\rm QCD}}\ ,\nonumber\\
&\Lambda_{\rm HS} \lesssim \Lambda_{\rm QCD}\qquad ~~~~~~~~~~~~~~~~~~ N\gtrsim \frac{M_{\rm pl}}{\Lambda_{\rm QCD}}\ ,
\end{align}
where $0\le \gamma \le \frac{1}{2}$. The fractional power of $M_{\rm pl}/N\Lambda_{\rm QCD}$ in equation \eqref{eq:BoundOnStringScale} arises because of interference effects \cite{Kaplan:2020ldi}. However, the exact value of $\gamma$ depends on form-factors and the knowledge of the spectrum and cannot be fixed by our argument. Moreover, causality necessarily requires that the Feynman diagram \ref{fig:feynman} must be accompanied by an infinite tower of higher spin exchanges  with unbounded spin \cite{Camanho:2014apa,Afkhami-Jeddi:2018apj,Caron-Huot:2016icg}.


Therefore, at this stage we conclude that there are  two  ways the conflict between large $N$ QCD and  causality can  be resolved:
\begin{enumerate}
\item[(I)] Glueball states -- The scattering amplitude $\mathcal{A}(s,t)$ has additional t-channel poles exactly at  locations of other glueballs.
\item[(II)] Gravity states -- The scattering amplitude $\mathcal{A}(s,t)$ develops new t-channel poles that represent  higher spin states in the gravity sector. \label{gravity}
\end{enumerate}
Of course,  the lightest higher spin t-channel pole must obey the bound (\ref{eq:BoundOnStringScale}) in both cases. For the remainder of this Letter, we will elaborate on these resolutions and argue that both have surprising implications.

 {\it \bf (I) Causality Restoration by Glueball States. ---} At first sight, it seems counter-intuitive to imagine that the gauge sector will remedy causality violations that are due to the graviton exchange. After all, we can replace the external gravitons by any other particle which is not in the QCD sector and the same problem persists. However, more careful consideration supports this possibility.

When large $N$ QCD is coupled to gravity, the scaling of matrix elements of glueballs with the graviton $h_{\mu\nu}$ are, in general, given by
\begin{align}
\label{eq:GGTScaling}
 \langle G_1\cdots G_n  h_{1} \cdots h_m \rangle \sim \frac{1}{N^{n-2}M_{\rm pl}^m}\ .
\end{align}
Importantly, all matrix elements with a single glueball and one or more gravitons can be $N$-enhanced
\be\label{eq:mixing}
 \langle G_J h_{\mu\nu} \rangle \sim \frac{N}{M_{\rm pl}}\ , \quad \langle G_J h_{\mu_1\nu_1} h_{\mu_2\nu_2} \rangle \sim \frac{N}{M_{\rm pl}^2}\ , \quad \cdots
\ee  
unless they are tuned to be zero. The first matrix element represents  a kinetic mixing between glueballs and the graviton. Of course, it is non-vanishing only when the graviton is off-shell and away from the soft-limit. Moreover, Lorentz invariance only allows mixing between spin-2 glueballs and the graviton \cite{comment_mixing}. The second matrix element corresponds to the decay of a glueball into two gravitons. These interactions are not part of our usual thinking about QCD, however, they can restore causality in the large $N$ limit.  
 
The presence of an infinite set of $\langle G_J h_{\mu_1\nu_1} h_{\mu_2\nu_2} \rangle$ immediately implies that  the scattering amplitude $\mathcal{A}(s,t)$, at the order $1/M_{\rm pl}^2$, has additional t-channel poles at  locations of other glueballs. Since glueballs have mass of parametric order $\Lambda_{\text{QCD}}$, these glueball poles obey the bound (\ref{eq:BoundOnStringScale}) implying that they can make $\mathcal{A}(s,t)$ consistent with causality.
 
On the other hand, the kinetic mixing $m_{\mu_1\mu_2}^{\nu_1\nu_2}(p)\equiv\langle G_{J=2}^{\nu_1\nu_2}(p) h_{\mu_1\mu_2}(-p) \rangle$ is already required to reproduce the 2-point function of the stress tensor \cite{juan}. In fact, one can derive a useful sum rule for such couplings
 \begin{align}\label{sumrule}
\langle T_{\mu_1\mu_2}(p)&T_{\mu_3 \mu_4}(-p)\rangle \nonumber\\
=&\frac{M_{\rm pl}^2}{4}\sum_{J=2} \frac{m_{\mu_1\mu_2}^{\nu_1\nu_2}(p)m_{\mu_3\mu_4}^{\nu_3\nu_4}(p) \Pi_{\nu_1\nu_2,\nu_3\nu_4}(p)}{p^2+m_J^2}
 \end{align}
 where the sum is over all spin-2 glueballs and $\Pi_{\nu_1\nu_2,\nu_3\nu_4}$ is the usual orthogonal projector for spin-2 exchanges. Note that the above sum rule implies that  an infinite number of such mixing must be non-zero in order to produce a 2-point function of the stress tensor which is consistent with the asymptotic freedom. Furthermore, in the presence of other non-QCD sectors, these kinetic mixing is also necessary in order to make large $N$ QCD causal  without modifying the gravity sector. For example, the non-QCD sector could just be a free particle with only gravitational interactions. One can replace the external gravitons in figure \ref{fig:feynman} by any other particle $X$ (with or without spin) and extend the argument of \cite{Kaplan:2020ldi} to conclude that the corresponding amplitude $G_J X \rightarrow G_J X$ must obey the same bound (\ref{eq:BoundOnStringScale}). However, now the causality violation can be avoided  if there is a  t-channel pole below $\Lambda_{\rm HS}$ with spin $J\ge 2$ \cite{comment_spin2}. The kinetic mixing term of (\ref{eq:mixing}) generates effective on-shell amplitudes $\langle XX G_{J=2}\rangle \sim N/M_{\rm pl}^2$ for all $X$. The effective interactions $\langle XX G_{J=2}\rangle$, in principle, can make the  amplitude $G_J X \rightarrow G_J X$ causal. Of course, causality  is a more precise constraint, beyond the sum rule (\ref{sumrule}). Furthermore, the graviton-glueball kinetic mixing is also required in order to generate effective three-point amplitudes of glueballs
 \be
 \langle G_{J}G'_{J'} G_{J=2}\rangle \sim \frac{N}{M_{\rm pl}^2}\ .
 \ee
In contrast to the canonical large $N$ scaling (\ref{eq:GGG}), this contribution is enhanced at large $N$. This  $N$-enhancement can, in principle, make the amplitude $G_J G'_{J'} \rightarrow G_J G'_{J'}$ consistent with causality even for  $N\gtrsim \frac{M_{\rm pl}}{\Lambda_{\rm QCD}}$.

 One might think of criticizing this solution on the grounds that massive spin-2 interactions are known to be highly constrained \cite{Camanho:2014apa,Bonifacio:2017nnt,Aragone:1979bm,Buchbinder:1999ar,Buchbinder:1999be,Buchbinder:2000fy,Zinoviev:2006im,Arkani-Hamed:2017jhn,Bonifacio:2017iry, Bonifacio:2018aon,Klaewer:2018yxi,deRham:2018dqm}. Moreover, in certain situations, as discussed in \cite{Camanho:2014apa}, they can introduce additional causality violations.  Thus, we do not know whether, and in what way, including a sum of spin-2 particles in the t-channel, can definitely save causality. We wish to emphasize that a better understanding of causality constraints when the graviton exchange is accompanied by a spin-2 tower would have important implications for large $N$ QCD. 
 
 So, the causality violations caused by the graviton can be resolved by the glueball states only if the IR interactions  (\ref{eq:mixing}) of glueballs take a  specific form \cite{comment_nogo}. In this scenario, the coefficients of higher dimensional operators (\ref{eq:mixing}) in the IR effective theory are tightly constrained by  causality. From the UV perspective, this IR `fine tuning' is unavoidable (but may be automatic from the UV Lagrangian). Whereas, from the IR perspective, it may appear that there has been a miraculous cancellation between the gravity sector and the QCD sector. Any low-energy terms that change these higher dimensional operators necessarily require new states in the gravity sector.

 {\it Spectating particles and classical shockwaves. ---} The non-trivial mixing between the gauge sector and the gravity sector due to (\ref{eq:mixing}) cannot fix causality violation when  spectating particles (with or without spin) are present in the theory. Because of the kinetic mixing, the meaning of ``spectating" particles is not obvious, so let us first give it a definite meaning. The scattering amplitude $\psi G_J \rightarrow \psi G_J$ of a spectating particle $\psi$ has t-channel poles only at the location of the graviton and other particles in the gravity sector (if any). Equivalently, spectating particles can be defined as particles that can be used to create classical (or stringy) gravitational shockwaves. Of course, when mixing terms of (\ref{eq:mixing}) are present, spectating particles are not the same as free particles. Furthermore, the preceding discussion implies that spectating particles, if present, lead to causality violations that can not be resolved by glueball states. Hence, spectating particles are ruled out unless we introduce new states in the gravity sector.

 {\it Free massless spin $\frac{3}{2}$ particles are ruled out. ---} We now consider the scattering process $G_J X_{3/2}\rightarrow G_J X_{3/2}$, where $X_{3/2}$ is a free massless spin $\frac{3}{2}$ particle \cite{comment_free}. We can  define a $2\times 2$ phase-shift matrix $\delta_{\pm,\pm}$, where $+$ and $-$ represent two helicities of the incoming and outgoing $X_{3/2}$. Causality implies that $\delta$ must be positive semi-definite which necessarily requires $\delta_{++}, \delta_{--}\ge 0$. An argument similar to the Weinberg-Witten theorem \cite{Weinberg:1980kq} ensures that the effective couplings $\langle X_{+}X_{+}G_{J=2}\rangle=\langle X_{-}X_{-}G_{J=2}\rangle=0$ because of angular momentum conservation, even when the mixing terms of (\ref{eq:mixing}) are present \cite{comment_3/2}. Moreover, other higher spin ($J\ge 3$) glueballs do not contribute as well since particle $X_{3/2}$ is  free. This implies that  glueball states cannot make the scattering process $G_J X_{3/2}\rightarrow G_J X_{3/2}$ causal. Thus, any theory of large $N$ QCD along with free massless spin $\frac{3}{2}$ particles is inconsistent with causality unless there exist higher spin states in the gravity sector obeying (\ref{eq:BoundOnStringScale}).

 {\it The regime $N\gtrsim \frac{M_{\rm pl}}{\Lambda_{\rm QCD}}$. ---} It follows from  (\ref{eq:mixing}) that  glueballs and mesons, in the presence of gravity, become more and more unstable as we increase $N$ keeping $M_{\rm pl}/\Lambda_{\rm QCD}$ fixed. Furthermore, as $N\sim \frac{M_{\rm pl}}{\Lambda_{\rm QCD}}$, the mixing between the QCD and the gravity sectors become significantly large. In particular, consider the scattering process $hh \rightarrow hh$ which now receives contributions at the order $1/M_{\rm pl}^2$ from higher spin glueball exchanges  because of interactions (\ref{eq:mixing}). If gravity is still weakly coupled, the theorem of \cite{Caron-Huot:2016icg}  implies that the resulting theory has a consistent S-matrix only if glueball states organize themselves into Regge trajectories that asymptotically coincide with the tree-level string theory spectrum, where the effective string scale is given by $\Lambda_{\rm QCD}$ \cite{Kaplan:2020ldi} \cite{comment_cksz}. Thus, the gravity-sector of the theory, if weakly coupled, has a natural description in terms of strings for $N\gtrsim \frac{M_{\rm pl}}{\Lambda_{\rm QCD}}$. Moreover, in this regime, it is tempting to interpret glueballs  as effective excitations of closed strings \cite{comment_closed}. 
 
{\it \bf  (II) Causality Restoration by Gravity States. ---} We might instead resolve the causality problem caused by the graviton by modifying the gravity sector. In fact, there are some plausible-seeming situations where it is the only way to restore causality:
 \begin{enumerate}
 \item[(i)] Phenomenologically one may choose to add higher dimensional operators in the IR effective theory that modify some of the interactions (\ref{eq:mixing}). 
 \item[(ii)] Interactions (\ref{eq:mixing}) are present but do not actually solve the causality problem by itself.  
 \item[(iii)] Spectating particles (whose scattering amplitudes only have gravity-sector  poles) are present. \label{item3}
 \item[(iv)] A free massless  spin $\frac{3}{2}$ particle is present. 
 \end{enumerate}
Again consider  the scattering amplitude $\mathcal{A}(s,t)$ of the process $G_J h_{\mu\nu}\rightarrow G_J h_{\mu\nu}$. Now, the causality violation due to the graviton pole is fixed by new t-channel poles that should be regarded as  higher spin  states in the gravity sector. Hence, metastable higher-spin glueballs (or mesons) in $3+1$ dimensions can couple to gravity while preserving causality if there exist higher spin states in the gravitational sector well below the Planck scale $M_{\rm pl}$. Furthermore, inequalities (\ref{eq:BoundOnStringScale}) now impose a  bound on the mass $\Lambda_{\rm HS}$ of the lightest higher spin particle in the gravity sector.

 The existence of these new gravitational states  implies that for $N\gtrsim \frac{M_{\rm pl}}{\Lambda_{\rm QCD}}$ the QFT approximation breaks down at the scale of $\Lambda_{\rm QCD}$ and hence even in this scenario we do not have a QCD theory in the traditional sense. Note that $N\gtrsim \frac{M_{\rm pl}}{\Lambda_{\rm QCD}}$ is precisely the regime in which  the gravitational interaction between glueballs is no longer weaker than the gauge interaction \cite{comment_wg}.

{\it UV completion. ---} This resolution has profound implications that stem from the fact that theories with higher spin exchanges are highly constrained by S-matrix consistency conditions \cite{Camanho:2014apa,Caron-Huot:2016icg}. First of all, any four-point amplitude with a finite number of higher spin exchanges is inconsistent with causality since   a spin-$J$ exchange produces a phase shift $\delta\sim s^{J-1}$. Thus, in this scenario, the gravity sector necessarily contains the graviton and an infinite tower of higher spin particles above $\Lambda_{\rm HS}$ with unbounded spin \cite{comment_loophole}. Furthermore, we assume that the gravity-sector of the underlying UV complete theory is weakly coupled and has a healthy thermodynamic limit.  This implies that the gravity spectrum does not have an accumulation point and the gravitational scattering amplitude  is a meromorphic function with simple poles obeying unitarity, causality, and crossing symmetry. Now invoking the theorems of \cite{Caron-Huot:2016icg,Kaplan:2020ldi} we conclude that the full gravity sector must contain infinite towers of  Regge trajectories that are 
asymptotically parallel, linear, and equispaced. In particular, the gravitational scattering amplitude in the regime $t,s\rightarrow \infty$ (such that  all intermediate scales  decouple) must coincide with the tree-level  four-point amplitude of fundamental closed strings \cite{Caron-Huot:2016icg,Kaplan:2020ldi,Gross:1987kza}
\be\label{amp_asym}
\lim_{s,t\gg 1}A_{\text gravity}(s,t)=A_0 e^{-\frac{\alpha'}{2}\(u\ln u+s\ln s+t\ln t\)}\ ,
\ee
where the Regge slope is $\alpha'\approx \frac{1}{\Lambda_{\rm HS}^2}$. The above amplitude has the feature that the inelastic part even for large impact parameter is  non-zero and consistent with the production of long strings.  Thus, in this scenario $\Lambda_{\rm HS}$ should be identified as the string scale $M_{\rm string}\approx \Lambda_{\rm HS}$. This is perfectly consistent with the fact that infinite towers of higher spin states in string theory have a well behaved S-matrix which respects asymptotic causality \cite{DAppollonio:2015fly,Amati:1987wq,Amati:1987uf,Amati:1988tn,Amati:1992zb,Amati:1993tb}.

\begin{figure}
\begin{center}
\includegraphics[width=0.49\textwidth]{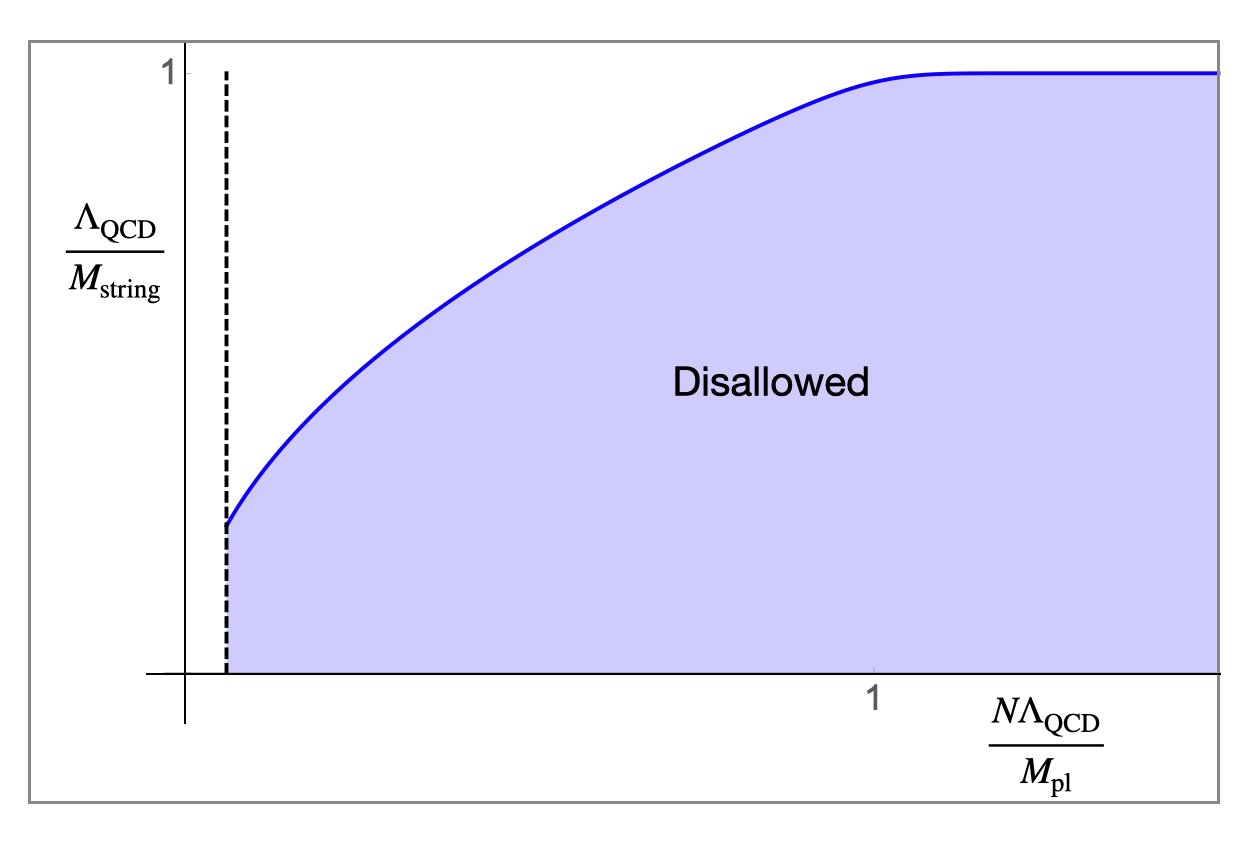}
\end{center}
\caption{ \small  A schematic exclusion plot for the string scale $M_{\rm string}\approx\Lambda_{\rm HS}$ as a function of $N$ when gravity states are restoring causality. The solid blue line represents the bound (\ref{eq:BoundOnStringScale}) for $\gamma=1/2$, where the shaded region is ruled out by causality. For $\gamma<1/2$, the bound asymptotes to 1 at a faster rate.  The dashed black line corresponds to an (unknown) large but finite value of $N$ above which our bounds hold. \label{fig:intro}}
\end{figure}

Therefore, when gravity states restore causality  of a confining large $N$ gauge theory, any weakly coupled UV completion of the resulting theory must have a gravity sector with many of the properties of fundamental strings \cite{comment_M}. The bound (\ref{eq:BoundOnStringScale}), in the present context, has a natural interpretation as a bound on the string scale. The claimed  bound on the string scale, as summarized in figure \ref{fig:intro}, appears quite surprising. For example, even a conservative estimate of (\ref{eq:BoundOnStringScale}) implies a parametric bound 
\be\label{eq:discussion}
M_{\rm string}\lesssim \sqrt{M_{\rm pl}\Lambda_{\rm QCD}/N}\quad \text{when}\quad  N\lesssim \frac{M_{\rm pl}}{\Lambda_{\rm QCD}}
\ee
but sufficiently large. On the other hand, $M_{\rm string}$ must be at or below  $\Lambda_{\rm QCD}$ for $N\gtrsim \frac{M_{\rm pl}}{\Lambda_{\rm QCD}}$, where scenarios (I) and (II) become practically indistinguishable.  

The above predictions are rather surprising, so one may try to find counter-examples to conclusively rule out the scenario (II). At first sight, the bound (\ref{eq:discussion}) seems to be in tension with the heterotic string theory. For example, one can start with  heterotic strings in 10d and study  compactification to 4d Minkowski space with $\mathcal{N}=2$ supersymmetry \cite{Kachru:1995wm}. In this construction,  $\Lambda_{\rm QCD}/M_{\rm string}$ is non-perturbatively small in the string coupling and hence appears to violate (\ref{eq:discussion}). However, these types of constructions can only provide a confining gauge theory coupled to gravity with $N$ up to order 10. There are other string constructions with similar features that can lead to slightly higher $N$, see \cite{Kachru:1995fv} for example. It is not actually a contradiction because our bound is applicable only above some large but finite $N$, where the $N$-cutoff is theory dependent but should not depend on various scales such as $M_{\rm pl}$, $\Lambda_{\rm QCD}$ or $\Lambda_{\rm HS}$. Unfortunately, our argument is too general to say anything more about what $N$ is large enough for our bound to apply. Nevertheless, we are not aware of any construction where $N$ is arbitrarily large providing a concrete counter-example.

At this stage, one may propose a natural question: what's the largest gauge group for a confining gauge theory coupled to gravity obtained from an explicit string construction?

{\it \bf Conclusions.---} Metastable higher-spin glueballs in four spacetime dimensions are only consistent with causality  when other higher spin states contribute to gravitational scattering amplitudes \cite{comment_caveat} \cite{comment_caveat2}. We have argued for an upper  bound on the mass $\Lambda_{\rm HS}$ of the lightest higher spin state required for the preservation of causality. These higher spin states can come from the glueball sector because of a non-trivial mixing between QCD and gravity. This rules out  the existence of  spectating particles that create classical shockwaves and massless free spin-$\frac{3}{2}$ particles. Causality can also be restored by a stringy gravity sector in which case we obtain a surprisingly strong bound on the string scale. A unifying feature of both these resolutions is that for large $N\gtrsim \frac{M_{\rm pl}}{\Lambda_{\rm QCD}}$ the gravity-sector of the resulting theory, if weakly coupled, only has a stringy description with the string scale at or below $\Lambda_{\rm QCD}$.  It would be valuable to improve our understanding of causality constraints involving towers of spin 2 particles, as this could impose additional constraints on glueball-graviton interactions.

\bigskip

\begin{acknowledgments}
It is our pleasure to thank Nima Afkhami-Jeddi,  Simon Caron-Huot, Liam Fitzpatrick, Shamit Kachru, David Kaplan,  Ami Katz, Juan Maldacena, and Amirhossein Tajdini for several helpful discussions.  We would also like to thank Simon Caron-Huot and Juan Maldacena for comments on a draft. We were supported in part by the Simons Collaboration Grant on the Non-Perturbative Bootstrap. JK was supported in part by NSF grant PHY-1454083.
\end{acknowledgments}

%


\begin{thebibliography}{102}
\bibitem{comment_meson}
 The same conclusion holds even for mesons which leads to a weaker bound. The
  bound for mesons can be obtained from (\ref{eq:BoundOnStringScale}) by
  replacing $N$ by $\sqrt{N}$.
\bibitem{comment_mixing}
  The only non-vanishing kinetic mixing term that can be written has the structure $h_{\mu\nu}\Box^k G_{J=2}^{\mu\nu}$.
\bibitem{comment_nogo}
It is also possible that interactions (\ref{eq:mixing}), however fine tuned, cannot actually solve the problem by itself.
\bibitem{comment_free}
To be precise, by massless free particle we mean the Lagrangian only has a kinetic term.
\bibitem{comment_M}
 Let us emphasize that this conclusion is valid only when the UV completion is weakly
  coupled. In fact, there are M-theory compactifications on manifolds with
  singularities that give rise to confining large $N$ gauge theories which have
  strongly coupled UV completion.
\bibitem{comment_caveat}
 To be clear, if large $N$ confining theories somehow do not contain higher
  spin glueballs in their spectra, none of our arguments would apply.
\bibitem{comment_caveat2}
 It is possible that a combination of scenario (I) and (II) restores causality in such a way that the bound \eqref{eq:discussion} can be relaxed.
\end{thebibliography}
\end{document}